\documentclass[12pt]{iopart}

\usepackage{iopams}

\def\be{\begin{equation}}
\def\ee{\end{equation}}
\def\ba{\begin{eqnarray}}
\def\ea{\end{eqnarray}}

\def\bra{\langle}
\def\ket{\rangle}

\def\o{\mathring{\omega}}
\def\e{\mathring{e}}
\def\q{\mathring{q}}
\def\vzero{\mathring{V}}
\def\SU(2){\rm SU(2)}

\def\t{\tilde}
\def\h{\hat}
\def\f{\frac}
\def\muv{\vec{\mu}}

\def\A{\mathfrak{A}}
\def\Abar{\bar\mathcal{A}}

\def\W{\mathfrak{W}}
\def\K{\mathcal{K}}
\def\V{\mathcal{V}}
\def\Diff{\mathit{Diff}}

\def\R{\mathbb{R}}
\def\C{\mathcal{C}}
\def\F{\mathbf{F}}
\def\lp{{\ell}_{\rm Pl}}

\def\H{{\cal H}}
\def\Hk{\H_{\rm kin}}

\def\muv{\vec{\mu}}
\def\etav{\vec{\eta}}

\begin{document}

\title{On the Uniqueness of Kinematics of Loop Quantum Cosmology }

\author{Abhay Ashtekar and Miguel Campiglia}
\address{Institute for Gravitation \& the Cosmos, and
    Physics Department, Penn State, University Park, PA 16802,
    USA}
\ead{ashtekar@gravity.psu.edu}

\begin{abstract}

The holonomy-flux algebra $\A$ of loop quantum gravity is known to
admit a natural representation that is uniquely singled out by the
requirement of covariance under spatial diffeomorphisms. In the
cosmological context, the requirement of spatial homogeneity
naturally reduces $\A$ to a much smaller algebra, $\A_{\rm Red}$,
used in loop quantum cosmology. In Bianchi I models, it is shown
that the requirement of covariance under \emph{residual}
diffeomorphism symmetries again uniquely selects the representation
of $\A_{\rm Red}$ that has been commonly used. We discuss the close
parallel between the two uniqueness results and also point out a
difference.

\end{abstract}


%\submitto{\CQG}
\maketitle

\section{Setting the Stage}
\label{s1}

In loop quantum gravity (LQG) one begins with a Hamiltonian
framework in which the basic canonical pair %$(A_a^i, \Sigma_{ab}^i)$
consists of an $\SU(2)$ connection $A_a^i$ and its momentum, a
Lie algebra-valued vector density $E^a_i$ of weight one, both
defined on a 3-dimensional manifold $M$. To construct quantum
kinematics, as usual, one has to select a class of elementary
functions which are to have unambiguous quantum analogs. In LQG
these are given by matrix elements of holonomies $h_\alpha(A)$
of connections $A$ along suitable curves $\alpha$ in $M$ and
fluxes $E_{S, f}$ of $E$ across suitable 2-surfaces $S$,
smeared with Lie algebra valued fields $f_i$. The kinematical
algebra $\A$
---called the \emph{holonomy flux algebra}--- is then generated by the
operators $\h{h}_\alpha$ and $\h{E}_{S,f}$ \cite{acz}. The
algebra $\A$ is `background independent' in the sense that it
uses only the manifold structure of $M$. To complete the
construction of quantum kinematics, then, it remains to find a
suitable Hilbert space $\Hk$ and represent elements of $\A$ by
concrete operators on it. Motivated by background independence,
$\Hk$ was taken to be the space $L^2(\Abar,\, {\rm d}\mu_o)$ of
square-integrable functions on the space $\Abar$ of (suitably
generalized) connections on $M$ with respect to a natural
\emph{diffeomorphism invariant measure} $\mu_o$ \cite{al2,jb1}.
The configuration operators $\h{h}_\gamma$ were represented by
multiplication and the momentum operators $\h{E}_{S,f}$ by Lie
derivatives w.r.t. certain `vector fields' on $\Abar$. This
representation of $\A$ admits a cyclic vector $\Psi_o$ which is
is invariant under the action of $\Diff$, the group of suitable
diffeomorphisms of $M$ \cite{al2}. This kinematics was
constructed in the mid nineties and led to a specific quantum
Riemannian geometry that underlies LQG \cite{alrev}.

However, a natural question arose: Is this representation of $\A$
\emph{uniquely} selected by some physical requirements? This was
answered in the affirmative some 10 years later through a powerful
theorem \cite{lost}: The physical requirement is precisely the
existence of a cyclic state invariant under $\Diff$,  which in turn
implies that the group $\Diff$ of symmetries is unitarily
implemented on $\Hk$. (See also \cite{cf}). This unitary
implementation plays a crucial role in the subsequent imposition of
the diffeomorphism constraint \cite{alrev}.

Let us now turn to cosmology. In the Bianchi I models we will
focus on, spatial homogeneity causes a drastic reduction in the
number of degrees of freedom. To obtain a simple description of
those that survive, one commonly introduces and fixes some
fiducial structures: a flat metric $\q_{ab}$, an associated set
of Cartesian coordinates $x^i$ on $M$, the associated
orthonormal co-frames $\o^i := dx^i$ and the dual frames
$\e^a_i$. One then restricts oneself to pairs $(A_a^i, E^a_j)$
of the form:%
\footnote{Throughout this communication, there is no summation
over repeated contravariant or covariant indices. Contracted
covariant and contravariant indices by contrast are summed
over. Our $c^i, p_j$ have been generally denoted by $\t{c}^i,
\t{p}_j$ in the LQC literature.}
\be \label{cp}
 A_a^i = c^i\,\o_a^i, \quad\quad E^a_i = p_i\, \sqrt{\q}\,\e^a_i \ee
where $\q$ is the determinant of the fiducial metric $\q_{ab}$.
Thus, because of spatial homogeneity, there are only three, global
configuration degrees of freedom $c^i$, and three momenta $p_i$.
However, if one naively evaluates the symplectic structure of the
full theory for these \emph{homogeneous} $A_a^i, E^a_i$, it
diverges. Therefore, to obtain a well-defined phase space
formulation and subsequent quantum kinematics, one must introduce an
infra-red cutoff (to be removed at the end to obtain physical
results). This is done by introducing a cell $\C$ whose edges are
parallel to the fiducial $\e^a_i$. Then, the non-vanishing Poisson
brackets are given by $\{c^i,\, p_j\} \,\,=\,\, (8\pi \gamma
G/\vzero)\,\,\delta^i_j$, where $\vzero$ is the volume of the cell
$\C$ with respect to the fiducial metric $\q_{ab}$.

To construct quantum kinematics, one begins by noting that it
is natural to restrict the holonomy and flux phase space
functions using spatial homogeneity. For fluxes, it suffices to
choose the surfaces to be the three faces of the cell (and
smearing fields $f_i$ to be $f_i = n_a \e^a_i$ where $n_a$ is
the unit normal to the face with respect to $\q_{ab}$). Then,
the three flux functions $E_{S,f}$ turn out to be (multiples of
the) $p_i$. For holonomies $h_\alpha$, it suffices to choose
the curves $\alpha$ to be aligned with the three edges of the
cell and label them with numbers $\mu_i$, the lengths of the
(oriented) edges in units of the edge lengths of the cell.
%(The edges of the cell are oriented and the $\mu_i$ are positive if
%$\alpha$ is parallel to the edge of the cell and negative if it is
%anti-parallel).
Then, if $\alpha$ is along the $j$th edge, $h_\alpha = (\cos
(\mu^jc^j)/2){I} + 2(\sin (\mu^jc^j)/2) \tau^j$ where $\tau^j$
are the Pauli matrices and ${I}$ the unit matrix. Note that the
dependence on $c^j$ is completely captured by the functions
$e^{i\mu^jc^j}$, with $\mu^j \in \mathbb{R}$. To summarize,
then, spatial homogeneity naturally reduces the holonomy flux
algebra $\A$ to the much smaller, reduced algebra $\A_{\rm
Red}$, generated by the phase space functions $e^{i\mu^jc^j}$
and $p_i$ \cite{awe2,asrev}.

While the reduction from $\A$ to $\A_{\rm Red}$ is systematic,
the construction of the representation of $\A_{\rm Red}$ used
in LQC has not descended so directly from LQG. For, while in
full LQG the representation was uniquely selected by asking for
a cyclic state which is invariant under $\Diff$, it was
generally believed that the ansatz (\ref{cp}) freezes all
diffeomorphisms. Thus the key requirement that selected the
unique representation in LQG seemed to have disappeared in LQC
whence it seemed impossible to prove an uniqueness theorem
along the lines of \cite{lost,cf}. Instead, one `mimicked' the
form of the unique cyclic state of LQG in a precise sense to
obtain a cyclic state on $\A_{\rm Red}$ and used it to
construct the representation \cite{abl}. In particular, the LQC
Hilbert space $\Hk$ is again the space $L^2(\mathbb{R}^3_{\rm
Bohr}, {\rm d}\mu_o)$ of square integrable functions on the
space $\mathbb{R}^3_{\rm Bohr}$ of (generalized) homogeneous
Bianchi I connections with respect to a natural measure ${\rm
d}\mu_o$ thereon, the holonomy operators $\widehat{\exp i
\mu^jc^j}$ act by multiplications and the flux operators
$\h{p_i}$ by derivation \cite{awe2,asrev}.

But the question has remained: Can we systematically arrive at
this representation in LQC as was done in LQG in
\cite{lost,cf}? The goal of this communication is to answer it
in the affirmative. The key new observation is that \emph{the
ansatz (\ref{cp}) does not eliminate the diffeomorphism freedom
completely and the residual diffeomorphism freedom can be used
to select a cyclic state on $\A_{\rm Red}$ uniquely.} Not
surprisingly this is precisely the state that was arrived at by
`mimicking' full LQG.

\section{Residual diffeomorphism symmetries}
\label{s2}

We have fixed the fiducial fields $\q_{ab}, \e^a_i, \o_a^i$ on $M$
and the Bianchi type I phase space variables are the connections
$A_a^i$ and conjugate momenta $E^a_i$ of the form (\ref{cp}). The
question is: Are there diffeomorphisms on $M$ which preserve this
form and have non-trivial action on the coefficients $c^i, p_j$? To
preserve the form (\ref{cp}), the diffeomorphisms must map each of
the three $\o_a^i$ to a constant multiple of itself. Since $\o^i =
dx^i$, it follows that the most general vector field $\xi^a$
generating such a diffeomorphism is a linear combination of
anisotropic dilations and translations:
\be \xi^a = \lambda_1 x_1 \e^a_1 + \lambda_2 x_2 \e^a_2 +\lambda_3
x_3 \e^a_3 + k^i\e^a_i \ee
where $\lambda_i, k^i$ are real constants. The action of
translations $k^i\e^a_i$ leaves each of the $c^i, p_j$ invariant.
Therefore it is just the 3-dimensional Abelian group $G$ generated
by the three anisotropic dilations, $x^1 \mapsto e^{\lambda_1}x^1$,
etc, that has a nontrivial action on $c^i, p_j$:
\be \label{action}
 c^1 \mapsto e^{\lambda_1} c^1,  \quad p_1 \mapsto
 e^ {\lambda_2 +\lambda_3} p_1\quad \hbox{\rm and cyclic permutations}. \ee
Are these phase space symmetries? A trivial calculation shows
that while the vanishing Poisson brackets between the three
$c^i$ and those among the three $p_i$ are preserved, the
non-vanishing ones are preserved if an only if $\lambda_1+
\lambda_2+\lambda_3 =0$. This is precisely the 2-dimensional
group $G_o$ of volume preserving anisotropic dilations. In the
main part of this communication we will focus just on $G_o$.

\section{The Weyl Algebra}
\label{s3}

The holonomy flux algebra is generated by $U(\muv):=
\widehat{\exp i\mu_j c^j}$ and $\hat{p}_i$. As usual, since it
is mathematically more convenient to deal with (the bounded)
unitary operators rather than (the unbounded) self-adjoint
ones, let us exponentiate $p_i$ and set $V(\etav) := \exp
i\eta^j \h{p}_j$ with $\muv \in \mathbb{R}^3$ and work with the
pairs $U(\muv), V(\etav)$. However, in the final picture we
need $\hat{p}_j$ to be well defined self adjoint operators.
This is easily achieved by demanding that \emph{in the final
representation the operators $V(\etav)$ should be continuous in
the parameters $\etav$.} There is no such a priori requirement
on $U(\muv)$ because in full LQG there is no operator
corresponding to the connections; only holonomies are well
defined operators. The classical Poisson brackets dictate the
algebraic structure of these operators:
\ba &&U(\vec{\mu}_1) U(\vec{\mu}_2) = U(\vec{\mu}_1 + \vec{\mu}_2);
\quad V(\vec{\eta}_1) V(\vec{\eta}_2) = V(\vec{\eta}_1 +
\vec{\eta}_2);\nonumber\\
&&U(\vec{\mu})V(\vec{\eta}) =
e^{-ik{\vec{\mu}\cdot\vec{\eta}}}\, V(\vec{\eta})U(\vec{\mu}),
\quad\quad \hbox{where $k = 8\pi\gamma\lp^2/\vzero$}. \ea
It is often convenient to work with a combination
\be W(\muv,\etav) := e^{\f{ik}{2}\vec{\mu}\cdot\vec{\eta}}\,\,
U(\muv) V(\etav) \ee
called the \emph{Weyl operators} satisfying the following star
relations and product rule:
\ba &&[W(\muv, \etav)]^\star = W(-\muv, - \etav), \nonumber\\
&&W(\muv_1, \etav_1)\,W(\muv_2, \etav_2) =
e^{-\f{ik}{2}(\muv_1\cdot\etav_2 -\muv_2\cdot\etav_1)}\,\,
W(\muv_1+\muv_2 , \etav_1+ \etav_2)\, .\ea
Note that the vector space $\W$ generated by finite linear
combinations of Weyl operators is closed under both operations
and is a $\star$-algebra. This is the Weyl algebra for the
Bianchi I model, the symmetry reduced version of the algebra
used in \cite{cf} for LQG.

As in the full theory, it is convenient to use the Gel'fand,
Naimark, Segal (GNS) construction \cite{gns} to find its
representation. This requires us to choose a normalized
\emph{positive linear functional} (PLF) $\F$ on $\W$, i.e., a
linear map, $\F: \W \rightarrow \mathbb{C}$, from the Weyl
algebra to the set of complex numbers, such that: i)
$\F(a^\star a) \ge 0$ for all $a\in \W$; and ii) $F(\mathbb{I})
= 1$, where $\mathbb{I}$ is the identity element of $\W$. The
choice made in LQC \cite{abl,awe2,asrev},
\be \F(W(\muv,\etav)) \,=\, \delta_{\muv, \vec{0}},\quad
{\hbox{\rm and extends to $\W$ by linearity}}, \ee
mimics the PLF used in full LQG \cite{lost,cf}. Since $\F$ is
continuous in $\etav$, in the resulting GNS Hilbert space $\H$
the unitary operators representing $V(\etav)$ are continuous in
the parameters $\etav$, and are therefore generated by
self-adjoint operators $\h{p}_i$. Thus, we have a
representation of the reduced holonomy-flux algebra $\A_{\rm
Red}$. The Hilbert space $\H$ is often described in terms of
the orthonormal basis $|\muv\rangle$ of eigenvectors of
$\h{p}_j$. The action of the basic operators is given by:
\be U(\muv) |\muv_o\rangle = |\muv + \muv_o\rangle, \quad {\rm
and} \quad V(\etav) |\muv_o\rangle = e^{ik\etav\cdot\muv_o}\,
|\muv_o\rangle \, . \ee
We will now show that this representation is uniquely selected by
the requirement that the PLF be invariant under the action of the
group $G_o$ of volume preserving anisotropic dilations.

\section{Uniqueness of the representation: Direct method}
\label{s4}

Since the induced action of $G_o$ on the phase space preserves the
symplectic structure, it provides a 2 parameter family $\Lambda
(\vec\lambda)$ of automorphisms on the Weyl algebra:
\be \Lambda(\vec\lambda) [W(\muv, \etav)] =
W(e^{\lambda_1}\mu_1,e^{\lambda_2}\mu_2,e^{\lambda_3}\mu_3;\,\,\,
e^{\lambda_2\lambda_3}\eta_1, e^{\lambda_3\lambda_1}\eta_2,
e^{\lambda_1\lambda_2}\eta_3)\ee
where $\lambda_1+ \lambda_2 + \lambda_3 =0$. As in the
uniqueness theorems of LQG kinematics \cite{lost,cf}, we now
seek a PLF $\F$ on $\W$ which is invariant under these
automorphisms. The cyclic state in the resulting GNS
representation would then be invariant under these residual
diffeomorphism symmetries, whence they would be represented by
unitary transformations on the GNS Hilbert space \cite{gns}. In
addition we require that $\F(W(\muv,\etav))$ be continuous in
$\etav$ so that operators $\h{p}_i$ will be well-defined and
the GNS Hilbert space will also carry a representation of the
holonomy-flux algebra $\A_{\rm Red}$.

For notational simplicity, let us set $F(\muv;\etav) :=
\F(W(\vec{\mu}, \vec{\eta}))$. Then the two conditions imply in
particular that $F$ must satisfy: i) $F(\vec{0};\, \etav) =
F(\vec{0};\,\, e^{-\lambda_1}\eta_1, e^{-\lambda_2}\eta_2,
e^{-\lambda_3}\eta_3)$ for any $\lambda_i \in \R$ satisfying
$\lambda_1+ \lambda_2 + \lambda_3 =0$; and, ii) $F(\vec{0};
\etav)$ is continuous in $\etav$. In addition, the
normalization condition on $\F$ implies $F(\vec{0}; \vec{0})
=1$. It follows immediately that:
\be \label{F} F(\vec{0};\, \eta_1,0,0) = 1; \,\,\, F(\vec{0};\,
0,\eta_2,0) = 1;\,\,\, F(\vec{0};\, 0,0,\eta_3) = 1. \ee
We are now equipped to prove the main result.
\smallskip

\textbf{Theorem:} Let $\F$ be a normalized positive linear
functional on the Weyl algebra $\W$ satisfying (\ref{F}). Then
$\F(W(\muv,\etav)) = \delta_{\muv, \vec{0}}$.

\noindent Proof: Being a PLF, $\F$ satisfies:
\be |\F(a^\star b)|^2 \le \F(a^\star a)\, \F(b^\star b)\quad
{\hbox{\rm for all}}\,\, a, b \in \W \, . \ee
The key idea is to use this property with two different choices
of $a$ and $b$. Let $\etav_o$ be any $\etav$ which lies along
one of the three axes so that $F(\vec{0};\, \etav_o) =1$. Set
$b = V(\etav_o) - \mathbb{I}$. Then it is trivial to check that
$\F(b^\star b) =0$. Therefore $\F(a^\star b) =0$ for all $a\in
\W$. Now let $a= V(\etav)$ for an arbitrary $\etav$. Then we
have $0=\F(a^\star b) = F(\vec{0};\,\etav_o-\etav) -
F(\vec{0};\,-\etav)$. Since $F(\vec{0}; \, \etav_o) = 1$, it
follows that $F(\vec{0}; \, \etav) = 1$ for all $\etav$.

Now let $b= V(\etav) - \mathbb{I}$ for any $\etav \in \R^3$.
Since we have established that $F(\vec{0}; \, \etav) = 1$, we
again have $\F(b^\star b)=0$, whence $\F(a^\star b) =0 =
\F(b^\star a)$ for all $a\in \W$. Therefore
$\F(a(V(\eta)-\mathbb{I}))=0$ and $\F((V(\eta)-\mathbb{I})a)=0$
for all $a \in \W$. This implies
\be \F(a) = \F(a V(\etav)) = \F(V(\etav)a)\quad \forall a\in \W
\,\,\, \hbox{\rm and}\,\, \etav \in \R^3. \ee
Let us now set $a = U(\muv)$ for any $\muv\in \R^3$. Then, using
$W(\muv,\etav) = e^{\f{i}{2} k \muv\cdot \etav}\, U(\muv) V(\etav)$,
we obtain
\be F(\muv;\,\etav) = e^{\f{i}{2} k \muv\cdot \etav}\, F(\muv;\,
\vec{0}) = e^{-\f{i}{2} k \muv\cdot \etav} \, F(\muv;\, \vec{0})\ee
for all $\muv, \etav$. This implies $F(\muv;\,\etav) = 0$ if
$\mu\not=0$. But we have already established that $F(\vec{0};\,
\eta) =1$. Therefore we conclude $F(\muv;\,\etav) = \delta_{\muv,
\vec{0}}$. $\quad \Box$
\smallskip

Thus, the requirement that the PLF be invariant under the
automorphisms on $\W$ implementing the residual diffeomorphism
symmetries $G_o$ led us to a unique cyclic representation of
$\W$. Moreover, this is precisely the representation that has
been used in LQC. Note, incidently, that $G_o$ invariance was
used only to arrive at the conclusion that $F(\vec{0};\,
\etav_o) =1$ for all $\etav$ on the three axes in the
3-dimensional $\eta$-space. So, if another physical requirement
were to lead us to this condition, uniqueness will follow. We
will return to this point in section \ref{s6} in the discussion
of more general Bianchi models.

\section{Uniqueness of the representation: Conceptual underpinning}
\label{s5}

It is instructive to see an alternate proof of the second half of
the uniqueness theorem because it makes the conceptual underpinning
of the result and the parallel between the LQC and LQG
representations transparent, and because it could extend to more
general situations. We begin by assuming that, thanks to the
symmetry condition, the PFL we are seeking must satisfy
$\F(W(\vec{0},\etav)) =1$. Let us suppose that such a PLF exists and
let $\K$ denote the kernel of the PLF, i.e., the subspace of $\W$
defined by $\F(a^\star a) = 0$ for all $a \in \K$. The GNS
construction then yields a Hilbert space $\H$ which is the Cauchy
completion of the quotient $\W/\K$.

The cyclic state $|\Psi_o\ket \in \H$ is the equivalence class to
which the identity operator $\mathbb{I}$ belongs. Since
$\F(\mathbb{I})=1$, we have $\bra \Psi_o|\Psi_o\ket =1$. Set
$|\Psi_{\etav}\ket = V(\etav)|\Psi_o\ket$. Then,
$\bra\Psi_{\etav}|\Psi_{\etav}\ket =1$ and furthermore $\bra
\Psi_o|\Psi_{\etav}\ket = \bra \Psi_o| V(\etav) \Psi_o\ket =
F(\vec{0};\, \etav) =1$. Thus, $|\Psi_o\ket$ and $|\Psi_{\etav}\ket$
are unit vectors and their scalar product is 1. Therefore they must
coincide. Thus, $V(\etav)|\Psi_o\ket = |\Psi_o\ket$ for all $\etav$.

Next, set $|\Psi_{\muv}\ket := U(\muv) |\Psi_o\ket$. Then
\be V(\etav) |\Psi_{\muv}\ket = V(\etav) U(\muv)|\Psi_o\ket =
e^{ik\muv\cdot\etav}\, U(\muv) V(\etav) |\Psi_o\ket =
e^{ik\muv\cdot\etav}\,|\Psi_{\muv}\ket \, . \ee
Thus, for all $\muv, \etav$,\,\,  $|\Psi_{\muv}\ket$ is an
eigenvector of $V(\etav)$ with eigenvalue
$e^{ik\muv\cdot\etav}$. Therefore it follows that: i) If $\muv
\not= \muv^\prime$,\, $|\Psi_{\muv}\ket -
|\Psi_{\muv^\prime}\ket \not\in \K$ so for each $\muv\in \R^3$
there is a distinct ket $|\Psi_{\muv}\ket$; and, ii) $\bra
\Psi_{\muv}| \Psi_{\muv^\prime}\ket = \delta_{\muv,
\muv^\prime}$. Consider the vector space $\V:= \{\sum_{n=1}^N\,
K_n|\Psi_{\muv_n}\ket \}$ spanned by finite but otherwise
arbitrary linear combinations of $|\Psi_{\muv}\ket$. It
contains the cyclic state $|\Psi_o\ket$ and is left invariant
by the Weyl algebra $\W$. Therefore $\V = \W/\K$, and its
Cauchy completion is the GNS Hilbert space $\H$. Thus we have
explicitly constructed the GNS representation. By inspection,
$\F(W(\muv,\etav)) = \bra \Psi_o|
e^{\frac{ik}{2}\muv\cdot\etav} U(\muv)V(\etav)|\Psi_o\ket =
\delta_{\muv, \vec{0}}$. Furthermore by identifying kets
$|\Psi_{\muv}\ket$ with the kets $|\muv\ket$ of section
\ref{s3}, we obtain an explicit isomorphism between this GNS
representation and the one that has been used in LQC.

\section{Discussion}
\label{s6}

We began by noting that the ansatz (\ref{cp}) used in the
Bianchi I models does \emph{not} completely fix the
diffeomorphism freedom. There is a three parameter group $G$ of
anisotropic dilations that respects the ansatz but has
non-trivial action on the symmetry reduced phase space. However
it is only the 2 parameter subgroup $G_o$ of volume preserving
diffeomorphisms of $G$ that preserve the symplectic structure.
Therefore we focused on $G_o$. This $G_o$ is faithfully
represented by a group of automorphisms on the Weyl algebra
$\W$. As is usual in quantum mechanics and quantum field
theory, we then seek cyclic representations of $\W$. If we
demand, as in full LQG \cite{lost,cf}, that the required PLF on
$\W$ be invariant under the automorphisms induced by the
diffeomorphism symmetries, we are led to a unique
representation of $\W$. Moreover this is precisely the
representation that has been used in the LQC literature
\cite{abl,awe2}. Thus the situation in LQC has turned out to be
completely parallel to that in LQG: \emph{the representation is
uniquely selected by the residual diffeomorphism symmetries.}
In both cases the representation was first found and used
extensively and the uniqueness was established much later.
%
%\footnote{The situation was the same with the Fock representations
%in free field theories in Minkowski space, where uniqueness was
%established only in the mid 1950s.}
%

We conclude with a few remarks:\\
i) While there is a conceptual parallel between LQG and LQC,
there is also a difference. If the topology is $\R^3$, the
group $G_o$ of diffeomorphism we considered is included in the
group $\Diff$ used in LQG \cite{lost}. However, the LQG
uniqueness result would have held even if one had restricted
oneself to diffeomorphisms which are asymptotically identity.
The uniqueness theorem would have still picked the standard PLF
and one could have just checked at the end that the PLF is also
invariant under the action of $G_o$. This difference is
directly related to the fact that we are now working with
homogeneous fields which do not have local degrees of freedom.

ii) In more general Bianchi models with different spatial
topologies, the analog of $G_o$ may not exist. But the induced
automorphisms continue to exist and can be interpreted as
changes of the fiducial $\o_a^i, \e^a_i$. Demanding that the
PLF be invariant under them would again lead to a unique cyclic
representation of the Weyl algebra.

iii) What is the situation with elements of $G$ with
$\lambda_1+\lambda_2+\lambda_3 \not=0$ which are \emph{not} in
$G_o$? Because the induced action of these elements of $G$ does
not preserve the symplectic structure, they do not yield
automorphisms on all of $\W$. But they do induce automorphisms
on the two Abelian sub-algebras of $\W$ generated separately by
$U(\muv)$ and $V(\etav)$. Our PLF is invariant under them.

iv) The spatially flat, isotropic Hilbert space of LQC states
is naturally embedded in our Bianchi I Hilbert space $\H$. In
this sense, the uniqueness result naturally descends from the
Bianchi I to the k=0 Friedmann model. However, what if one
chose to work directly with the Friedmann model? Then,
$\lambda_1 = \lambda_2=\lambda_3$ and $G$ reduces just to the
1-parameter group of dilations. The action of this
diffeomorphism induces automorphisms only on the two Abelian
sub-algebras as discussed above. However, the requirement that
the PLF be invariant under this action suffices to select the
PLF uniquely \cite{ach4} and this is precisely the PLF that has
been used in the Friedmann model of LQC \cite{abl}. By contrast
in the Schr\"odinger representation, discussed below, this
1-parameter group of dilations is not unitarily implemented
and, furthermore, the Friedmann Hilbert space is not a subspace
of the Bianchi I Hilbert space.

v) What happens in the Schr\"odinger representation of the Weyl
algebra $\W$, where the Hilbert space is $L^2(\R^3, {\rm
d}^3c)$? Since $c^1 \mapsto e^{\lambda_1}\, c^1$, etc, with
$\lambda_1+ \lambda_2+\lambda_3 =0$, it follows that the
Lesbegue measure is preserved, whence $G_o$ is again unitarily
represented. Furthermore, this representation is again cyclic
but it does \emph{not} admit any cyclic state that is invariant
under the induced action of $G_o$!  This raises an interesting
question: Are there perhaps cyclic representations of the
holonomy-flux algebra $\A$ of LQG in which $\Diff$ is unitarily
represented but none of the cyclic vectors is invariant under
$\Diff$? \emph{If} they do, they could represent different
phases of LQG kinematics, complementing the standard
representation \cite{al2,alrev} which captures the LQG quantum
geometry at the Planck scale.

\textbf{Acknowledgements:} We would like to thank Adam Henderson for
discussions on the isotropic case and Jerzy Lewandowski and Hanno
Sahlmann for correspondence on the uniqueness result in LQG. This
work was supported in part by the NSF grant PHY-1205388, the Eberly
Research Funds of Penn State and a Frymoyer Fellowship.


\section*{References}
\begin{thebibliography}{99}

\bibitem{acz} Ashtekar  A and Isham  C J 1992 Representation of the
    holonomy algebras of gravity and non-Abelian gauge theories
    \textit{Class. Quant. Grav.} \textbf{9}  1433--1467;\\
    Ashtekar A, Corichi A and Zapata J A 1998 Quantum
    theory of geometry: III. Non-commutativity of Riemannian
    structures  \textit{ Class. Quant. Grav.} \textbf{15} 2955--2972
\bibitem{al2} Ashtekar  A and Lewandowski  J 1994
    Representation theory of analytic holonomy algebras, in
    \textit{Knots and Quantum Gravity}  ed Baez J C (Oxford U.\ Press,
    Oxford);\\
    1995 Differential geometry on the space of connections using
    projective techniques \textit{Jour. Geo. \& \ Phys.} \textbf{17}
    191--230
\bibitem{jb1} Baez  J C  1994  Generalized measures in gauge theory
    \textit{Lett. Math. Phys.} \textbf{31}  213--223
\bibitem{alrev} Ashtekar A and Lewandowski A, 2004 {Background
    independent quantum gravity: A status report}, Class. Quant.
    Grav. {\bf 21} R53-R152
\bibitem{lost} Lewandowski  J,  Oko\l\'ow  A, Sahlmann H,
    Thiemann T 2006 Uniqueness of the diffeomorphism invariant state
    on the quantum holonomy-flux algebras, Comm. Math. Phys. \textbf{267}
    703-733
\bibitem{cf} Fleishchack C, 2009 {Representations of the Weyl
    algebra in quantum geometry}, Commun. Math. Phys. \textbf{285} 67-140
\bibitem{abl} A.~Ashtekar, M.~Bojowald and J.~Lewandowski,
    {Mathematical structure of loop quantum cosmology}. Adv. Theo.
    Math. Phys. \textbf{7} 233--268 (2003)
\bibitem{gns} Gel'fand I M and Naimark M A, 1943 On the
    embedding of normed rings into the ring of operators in
    Hilbert space, Mat. Sobrn. \textbf{12, [54]} 197-217;  \\
Segal I E, 1947 Postulates of general quantum mechanics, Ann. Math.
\textbf{48} 930-948
\bibitem{awe2} Ashtekar A and Wilson-Ewing E, 2009 {Loop quantum
    cosmology of Bianchi type I models}, Phys. Rev. D\textbf{79} 083535
\bibitem{asrev} Ashtekar A and Singh P, 2009 Loop Quantum Cosmology:
    A Status Report, Class.\ Quant.\ Grav.\  {\bf 28}, 213001
\bibitem{ach4} Ashtekar A, Campiglia M and Henderson A 2011
    (unpublished notes)


\end{thebibliography}
\end{document}